\documentclass[usenatbib]{mn2e}

\usepackage{graphicx}
\usepackage{color}
\usepackage{amsfonts,amsmath,amssymb}

\newcommand{\ogle}{OGLE-TR-113b}

\begin{document}

\title[TraMoS IV: Timing Analysis of 20 transits of OGLE-TR-113b]{TraMoS IV: Discarding the Quick Orbital Decay Hypothesis for OGLE-TR-113b}

\author[S. Hoyer et al.] {  S.~Hoyer$^{1,2}$, M.~L\'opez-Morales$^{3}$, P.~Rojo$^{4}$, D.~Minniti$^{5,6,7,8}$, E.R.Adams$^{9}$ \\
$^1$ Instituto de Astrof{\'i}sica de Canarias, E-38205 La Laguna, Tenerife, Spain. \\
$^2$ Universidad de La Laguna, Dpto. Astrof{\'i}sica, E-38206 La Laguna, Tenerife, Spain.\\
$^3$ Harvard-Smithsonian Center for Astrophysics, 60 Garden Street, Cambridge, MA 01238, USA \\
$^4$ Universidad de Chile, Departamento de Astronom{\'i}a, Casilla 36-D, Santiago, Chile.\\
$^5$ Departamento de Ciencias F{\'i}sicas, Universidad Andr{\'e}s Bello, Rep{\'u}blica 220, Santiago, Chile.\\
$^6$Vatican Observatory, V00120 Vatican City State, Italy.\\
$^7$Millennium Institute for Astrophysics, Santiago, Chile.\\
$^8$Instituto de Astrof{\'i}sica, Pontificia Universidad Cat{\'o}lica de Chile, Av. Vicu{\~n}a Mackenna 4860, Santiago, Chile\\
$^9$Planetary Science Institute, 1700 East Fort Lowell, Suite 106, Tucson, AZ 85719 }

\maketitle

\begin{abstract}
In the context of the TraMoS project we present nine new transit
observations of the exoplanet OGLE-TR-113b observed with the Gemini
South, Magellan Baade, Danish-1.54m and SOAR telescopes. We perform a
homogeneous analysis of these new transits together with ten
literature transits to probe into the potential detection of an
orbital decay for this planet reported by \citet{adams2010}. Our new
observations extend the transit monitoring baseline for this system by
6 years, to a total of more than 13 years.  With our timing analysis
we obtained a $\dot{P}=-1.0 \pm 6.0$ ms~yr$^{-1}$, which rejects
previous hints of a larger orbital decay for OGLE-TR-113b.  With our
updated value of $\dot{P}$ we can discard tidal quality factors of
$Q_{\star} < 10^{5}$ for its host star.  Additionally, we calculate a
1$\sigma$ dispersion of the Transit Timing Variations (TTVs) of 42
seconds over the 13 years baseline, which discards additional planets
in the system more massive than $0.5-3.0~M_{\oplus}$ in 1:2, 5:3, 2:1
and 3:1 Mean Motion Resonances with OGLE-TR-113b.  Finally, with the
joint analysis of the 19 light curves we update transit parameters,
such as the relative semi-major axis $a / R_s = 6.44^{+0.04}_{-0.05}$, the
planet-to-star radius ratio $R_p / R_s =0.14436^{+0.00096}_{-0.00088}$, 
and constrains its orbital inclination to $i =89.27^{+0.51}_{-0.68}$~degrees.

\end{abstract}


\setlength{\tabcolsep}{0.2em} 

\section{Introduction}

OGLE-TR-113b was one of the first discovered transiting exoplanets,
reported by \citet{udalski2002} as a planet candidate orbiting a V =
16.1 K-dwarf star, and later confirmed by \citet{bouchy2004} and
\citet{Konacki_2004} via radial velocity follow-up campaigns. With a
mass of 1.23 $M_{\rm Jup}$ and a radius of 1.09 $R_{\rm Jup}$
\citep{southworth2012}, OGLE-TR-113b is a hot Jupiter orbiting its
host star once every 1.43 days. Due to the proximity to its host star,
OGLE-TR-113b is potentially an interesting target for orbital decay by
tidal dissipation studies \citep[see
  e.g. ][]{sasselov2003,patzold2004,carone2007,levrard2009,matsumura2010,penev2011,penev2012},
in which it is predicted that the orbital separation between the star
and the planet will continue to shrink -- in spite of orbital
circularization -- as long as the orbital motion of the planet is
faster than the stellar rotation rate. In those cases, the planet's
orbital decay will continue until the planet reaches the stellar Roche
radius limit of the system and falls into the star \citep[see
  e.g. ][]{levrard2009}.

Although the orbital decay of exoplanets is a topic that has received
increasing attention over the past decade, estimations of the expected
timescales of this effect remain largely unconstrained because of the
currently limited understanding and measurements of tidal dissipation
mechanisms in both planets and stars. Because of this lack of
understanding, tidal quality factors, which are a measure of the star
or planet's distorstion due to tidal effects and drive the efficiency
of the orbital time decay, are generally allowed to adopt a wide range
of values, between $Q_{\star} = 10^4$ -- $10^{10}$ \citep[see
  e.g. ][]{patzold2004,matsumura2010}.

Directly measuring the orbital decay of a close-in, short period,
exoplanet would enable the first empirical test to current tidal
stability and dynamical models of these objects. A way to detect that
orbital decay is via long-term monitoring campaigns of transiting
exoplanets in search for small and steady transit timing variations
\citep[TTVs; see e.g. ][]{miralda2002,holman2005,agol2005}, which
would show the transits occurring systematically closer in time over
timescales of several years.

\citet{adams2010}, hereafter A10, reported the tentative detection of
an orbital period decay of $\dot{P}=-60\pm 15$ ms~yr$^{-1}$ for
OGLE-TR-113b, but the authors acknowledged that more observations were
needed to confirm their claim. That period decay rate could be
reproduced by a relatively small tidal quality factor for the star of
$Q_{\star} \sim 10^{3} - 10^{4}$ \citep{Birkby2014}, which is close to
the theoretical lowest estimate for this parameter. Additionally,
\citet{penev2012} concluded that the population of currently known
planets is inconsistent at the 99$\%$ level with $Q_{\star} > 10^7$.

OGLE-TR-113b is one of the targets we have been monitoring in our
Transit Monitoring in the South (TraMoS) project, which includes
observations from the 1-m telescope at CTIO, SOAR and Gemini South
telescopes at Cerro Pach\'on Observatory \citep{hoyer2012}. TraMoS,
which has been underway since 2008, is dedidacted to searching for
transit timing variations of known planets to unveil additional
planets in those systems and, therefore, their architecture. Other
planetary systems we have published as part of TraMoS are OGLE-TR-111b
WASP-5b and WASP-4b \citep[see][]{hoyer2011, hoyer2012, hoyer2013}.

In this work we present eight new transit light curves of OGLE-TR-113b
from TraMoS, observed with Gemini South, SOAR and Danish-1.54m
telescopes, and a new transit light curve obtained with the same
instrumental setup used by A10 on Magellan. We combine those new light
curves with all available literature light curves to perform a new
study of transit timing variations for this system.  In Section
\ref{observaciones} we describe the observations. Section \ref{fiteo}
describes the data analysis and light curve fitting. Sections
\ref{analisis} and \ref{mercury} describe the timing analysis and mass
limits for unseen perturbers, and we present our conclusions in
Section \ref{conclusiones}.

\begin{table*}
\begin{tabular}{lcccccc}
\hline
\hline
Epoch & Date Obs. & Instrument /  & Filter & Average & Airmass range & \# points \\
& (yyyymmdd) & Telescope & & Cadence (s) & & \\
\hline
192   & 20060104 & GMOS/Gemini-S            & g'     & 75 / 330      & 2.34 - 1.18   & 120       \\
946   & 20081219 & GMOS/Gemini-S            & g'     & 54              & 2.07 - 1.29   & 180       \\
953   & 20081229 & GMOS/Gemini-S            & g'     & 54              & 1.67 - 1.21   & 206       \\
969   & 20090121 & GMOS/Gemini-S            & g'     & 54              & 1.65 - 1.20    & 213       \\
990   & 20090220 & GMOS/Gemini-S            & i'     & 44              & 1.19 - 1.31   & 284       \\
992   & 20090223 & GMOS/Gemini-S            & i'     & 49              & 1.58 - 1.17   & 381       \\
1471  & 20110110 & MagIC-e2V/Magellan         & i'     & 62              & 1.47 - 1.19  & 241      \\
2530  & 20150306 &  DFOSC/Danish-1.54m         & R     &  144           & 1.19 - 1.99  & 161      \\
2585 & 20150624 & SOI/SOAR         & I               & 57  & 1.18 - 1.92 &239    \\

\hline
\hline
\end{tabular}
\caption{\label{tab:observations} Description of each of the nine new transit observations presented in this work.}
\end{table*}

\section{Observations and Photometry} \label{observaciones}

We observed OGLE-TR-113b during nine transit epochs between 2006 and
2015. The first six transits were observed with the Gemini
Multi-Object Spectrograph (GMOS-S) instrument on the 8.1m Gemini South
Telescope (programs ID: GS-2005B-Q-9, GS-2008B-Q-11, GS-2009A-Q-16 and
GS-2010A-Q-36). GMOS-S in imaging mode has a pixel scale of 0.073
$arcsec/pixel$ and a Field-of-View (FoV) of 330$\times$300
$arcsec^2$. However, for these observations we used a Region of
Interest (RoI) which includes only the central 1024 rows, reducing the
readout time of the detector to only $\sim$47 seconds. The FoV of the
RoI is 75 $\times$ 168 $arcsec^2$, which given the relatively crowded
field of OGLE-TR-113b, contains enough comparison stars to perform
precise differential photometry. In addition, the high resolution of
the pixels minimizes blends.
  
The transit on 2006-01-04 ($E=192$, where we use as $E=0$ the transit
of 2005-04-04 from \cite{Gillon_2006} described below), was observed
alternating between the GMOS g'(G0325) and GMOS i' (G0327) filters
with exposures of 30 seconds each. Unfortunately, the GMOS i' images
were saturated and are not included in this work. The next three
transits were observed in the GMOS g'(G0325) filter and the last two
epochs were observed with the GMOS i' (G0327) filter. Each observation
lasted between 3.1 to 5.4 hours, and included the full transit and out
of eclipse baseline. The dates and other specific details of each
transit observation are summarized in Table \ref{tab:observations}.
 
 The transit of 2011-01-10 was observed with the MagIC-e2v camera on
 the 6.5m Baade Telescope at Las Campanas Observatory, and with the
 same setup described in A10. MagIC-e2v has a FoV of 38$\times$38
 $arcsec^2$, with a resolution of 0.037 $arcsec/pixel$. The frame
 transfer mode of MagIC-e2v provides a readout of 0.003 seconds per
 frame, which highly surpasses the readout of conventional cameras,
 such as GMOS-S. The observations were done in unbinned mode, with a
 Sloan i' filter, and an exposure time of 30 seconds per frame. The
 observations lasted 4.1 hours, and include the full transit and out
 of transit baseline.

The 2015-03-06 transit was observed using the DFOSC (Danish Faint
Object Spectrograph and Camera) camera on the 1.54m Danish Telescope
at ESO La Silla Observatory.  DFOSC has a field of view of
13.7'$\times$13.7' at a plate scale of 0.396 $arcsec/pixel$.  We used
unbinned mode, with the $Bessel ~R$ filter and an exposure time of 100
seconds.

The last transit, 2015-06-24, was obtained with SOI (SOAR Optical
Imager) on the 4.1m Southern Astrophysical Research (SOAR) Telescope
at Cerro Pach{\'o}n Observatory.  SOI is a mini-mosaic of two E2V
2k$\times$4k CCDs with a FoV of 5.26'$\times$5.26' and a pixel scale
of 0.077 $arcsec/pixel$.  We used a $Bessel ~I$ filter and an exposure
time of 45 seconds per frame in the 2x2 binned mode.  At the end of
the night the sky was covered by clouds which prevented observations
of the egress and after-the-transit baseline.

To reduce the data, in the case of GMOS-S we used the processed images
delivered by the Gemini telescope reduction pipeline. In the case of
the MagIC-e2v, DFOSC and SOI data, we bias-corrected and flatfielded
the images using standard IRAF routines.

The reduced images were ran through a custom, python-based
pipeline developed for TraMoS. This pipeline performs aperture
photometry of the target and a set of reference stars and combines
them to create differential light curves, free of most Earth
atmospheric effects. The aperture radius, sky annulus, and reference
stars are determined iteratively by the pipeline, as those that
produce the smallest dispersion of the out-of-transit light curves. In
some datasets, where the seeing variations during the night are large,
the pipeline allows for different values of the aperture and sky
annulus throughout the night. The light curves of each of
the nine new transits, which still contain some systematics effects
that need to be modelled (see Section \ref{fiteo}), are shown in
Figure \ref{fig:lcs} along with the literature light curves described
below.

\subsection{Literature Light Curves} \label{literature_lcs}
In our analysis we also included ten literature light curves: two
light curves in R band collected on April 4 and 14 2005 UT
\citep{Gillon_2006}, a V band light curve collected on April 11 2005
UT \citep{Pietrukowicz_2010,Diaz_2007}, a light curve in K band
observed on March 18 2006 UT \citep{Snellen_2007}, and six light
curves observed between January 30 2007 UT and May 10 2009 UT reported
by A10.  We used the compilation of all these light curves by A10.
\section{Light Curve Modelling} \label{fiteo}

We modelled our nine  new transit light curves simultaneously with the
literature light curves using the Transit Analysis Package \citep[TAP
  v2.104,][]{Gazak_2012} Like in A10, we did not fit the
\citet{Konacki_2004} light curve, since it is the result
of phase folded data from the OGLE survey over several transit epochs. 
Instead, we adopted their reported midtime of transit
and used that value in parts of the TTV analysis described in Section
\ref{analisis}.

We fit all the other light curves for the transits central time,
$T_c$, the planet-to-star radius ratio ($R_p/R_s$), the orbital
inclination ($i$), and a quadratic limb darkening law, with $u_1$ and
$u_2$ as the linear and quadratic limb darkening coefficients.  
We also fit a linear function of the flux vs time
($Y_{int}$ and $Y_{slope}$) in order to remove systematics in the
light curves, which are mostly produced by changes in the
airmass during the observations.  The amount of correlated and
uncorrelated noise is also estimated in each light curve using the
wavelet-based method proposed by \citet{Carter_2009}, where the noise
parameters, $\sigma_w$ (for the white noise) and $\sigma_r$ (for the
correlated noise), are fitted from the light curves assuming the
correlated noise follows a power spectral density varying as $1/f$.

We fixed the values of the orbital eccentricity, $e$, and longitude of
periastron, $\omega$, to zero, and we adopted a fixed orbital period for
the system of $P=1.43247425$~ days from A10.  

Having several transit epochs is advantageous to refine the values of
some of the system's parameters, such as $i$, $R_p/R_s$, and
$a/R_p$. Therefore, we fit for those parameters using all the light
curves, simultaneously, while letting $T_c$ vary individually for each
transit.  We found that we cannot produce reliable limb darkening
fits on individual light curves. The fits also had problems
distinguishing between very similar filters, e.g. between the Gemini
$i'$ and the MagIC $i'$ filters, or the Gemini $g'$ and $V$
filters. We got around this problem by fitting both limb darkening
coefficients ($u_1$ and $u_2$) simultaneously for all the
\textit{same} filter light curves, i.e. $i'$, $g'$, $R$ and $K$, where
we assumed that the limb darkening coefficients for similar filters
were the same.  Furthemore, based on \citet{Csizmadia_2012}, we do not
fix the limb darkening coefficients to theoretical predictions but leave 
them as free parameters.  
The limb darkening coefficients obtained from the joint analysis of
each filter are summarized in Table \ref{tab:resultados2}.

\begin{figure*}
\begin{center}
\includegraphics[width=1.8\columnwidth]{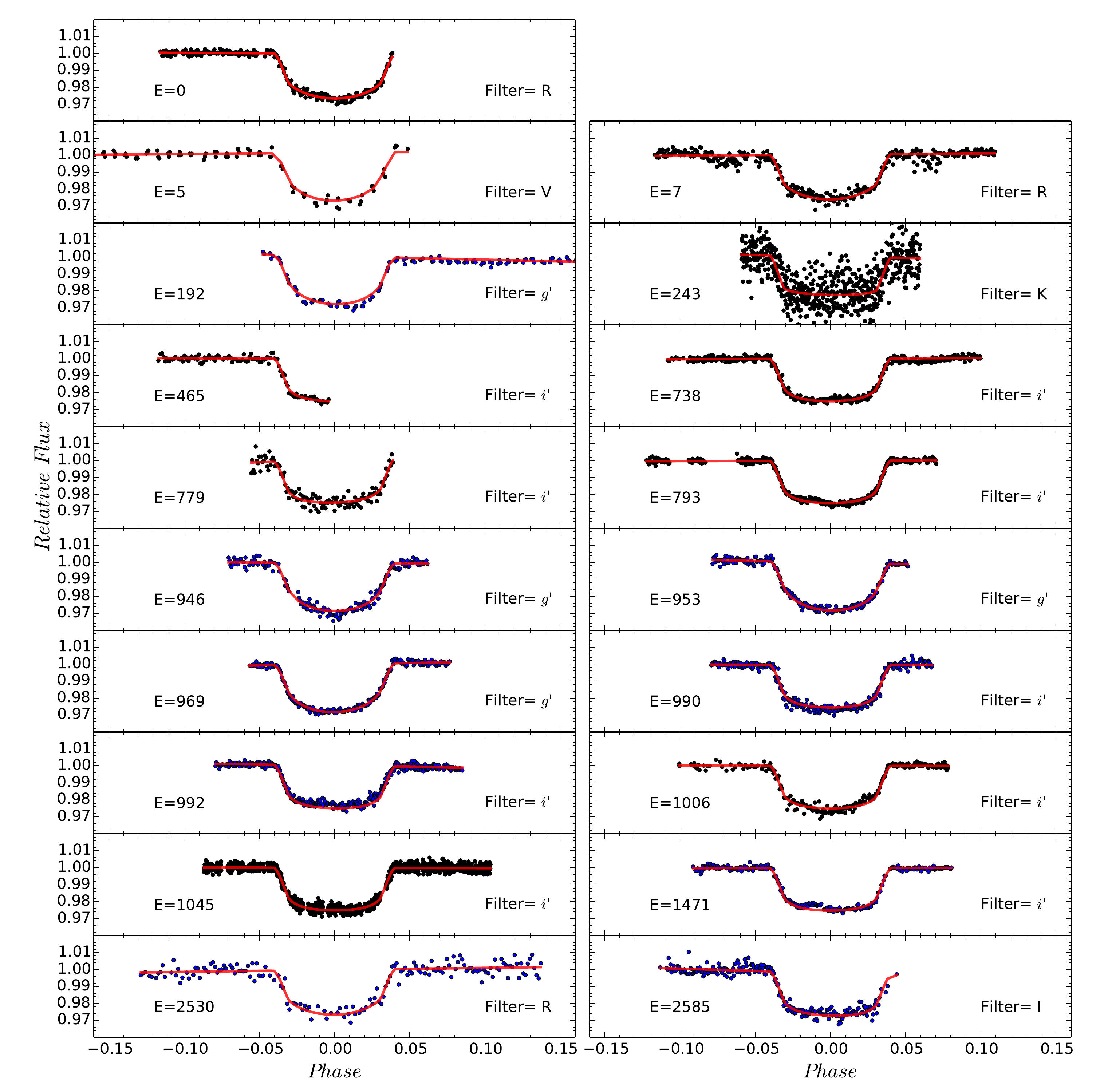}
\caption{\label{fig:lcs} Transit light curves of OGLE-TR-113b.  The
  nine transits presented in this work (blue points) are shown with
  the ten literature light curves (black points).  Each transit is
  labeled with its respective epoch and filter.  The transit models
  obtained with \textit{TAP} are shown with the red solid lines.}
\end{center}
\end{figure*}

We ran 10 different MCMC chains of $10^5$ links each, discarding the
first $10\%$ to avoid any bias introduced by initial values of the
fitted parameters.  Our fits yield refined values for $i$, $R_p/R_s$,
and $a/R_p$, which are summarized in Table \ref{tab:resultados2}. 
In Table \ref{tab:resultados1} we show the
central time obtained for each transit. The raw transit light curves
are shown in Figure \ref{fig:lcs}, together with their best model
fits. All the data are available online in tables including the times
and normalized fluxes of each transit; Table \ref{tab:LCS} shows an
excerpt of those tables.   

We note in the $E=1471$ light curve a signature that can be attributed
to star spot occultations of the planet during the transit.  The
transits $E=192$, $793$ and $992$ also show {\it bumps} in the light
curves during transit but with very low amplitudes.  Moreover, A10
reported that the bump in the $E = 793$ light curve was produced by
a rapid seeing variation. The large time span between these {\it
  detections} prevents us from carrying out a more detailed study of
the rotational period of the star.

\begin{table}
\begin{center}
\begin{tabular}{lcc}
\hline
\hline
\multicolumn{2}{l}{Simultaneous Fitted Parameters}&\\ 
\hline
\smallskip
\smallskip
$a/R_s $  & $6.44^{+0.04}_{-0.05}$  & \\ 
\smallskip
\smallskip
$(R_p/R_s )$ & $0.14436^{+0.00096}_{-0.00088}$& \\ 
$i$ (deg)    & $89.27^{+0.51}_{-0.68}$ &\\ 
\hline
Limb Darkening  & $u_1$    & $u_2$ \\
\hline
\smallskip
\smallskip
R &  $0.67^{+0.15}_{-0.18}$ &$ -0.03^{+0.29}_{-0.25}$\\
\smallskip
\smallskip
$g$', V & $0.733^{+0.094}_{-0.089}$ &$0.14^{+0.14}_{-0.15}$\\ 
\smallskip
\smallskip
K  & $0.20^{+0.25}_{-0.15}$ & $0.09^{+0.29}_{-0.33}$ \\
$i$' , I & $0.299^{+0.061}_{-0.055}$ & $0.43^{+0.10}_{-0.12}$\\
\hline
\hline
\end{tabular}
\end{center}

\caption{ \label{tab:resultados2} Results of the joint fit of the 19
  transits of \ogle.}  
\end{table}

\begin{table}
\begin{tabular}{lcccc}
\hline
\footnotesize{Epoch}  & \small{residuals}  &$T_{c} - 2450000.$  &  \small{ (O-C)} & \small{(O-C)}     \\  
 &\small{(ppm)}& \tiny(BJD$_{TDB}$) &  \tiny{lineal} (s) & \tiny{quad}~(s)                                                                                                   \\  
 \hline
\hline
-795  & -- & 2325.79897$^{+0.00082} _{- 0.00082}$ &                                             -38      &       -32    \\     
0  & $0.0013$ & 3464.61725 $^{+0.00027}_{-0.00026}$ &                                           15        &       16    \\     
5  & $0.0019$    &  3471.77859  $^{+ 0.00042  }_{- 0.00041}$ &                                  -75       &     -73    \\      
7& $0.0027$   &  3474.64382  $^{+ 0.00058  }_{- 0.00057}$ &                                     -51       &       -49   \\     
192 & $0.0017$ &  3739.65294  $^{+ 0.00052  }_{- 0.00053}$ &                                    56        &        57   \\     
243& $0.0087$ & 3812.70856  $^{+ 0.00060  }_{- 0.00061}$ &                                         4      &         5      \\  
465  & $0.0014$ & 4130.71840  $^{+ 0.00055  }_{- 0.00054}$ &                                    36         &      37    \\     
738 & $0.0013$  &  4521.78374  $^{+ 0.00023  }_{- 0.00023}$ &                                   6           &        6  \\     
779 & $0.0029$  &   4580.51525  $^{+ 0.00051  }_{- 0.00052}$ &                                  9         &         9   \\     
793 & $0.0010$  &  4600.56977  $^{+ 0.00014  }_{- 0.00014}$ &                                   -3         &      -3    \\     
946 & $0.0020$ &  4819.73961  $^{+ 0.00040  }_{- 0.00040}$ &                                      97       &       98   \\     
953 & $0.0014$  &  4829.76632  $^{+ 0.00030  }_{- 0.00032}$ &                                   44        &      44     \\     
969  & $0.0011$  &  4852.68574  $^{+ 0.00026  }_{- 0.00024}$ &                                  28      &          29   \\     
990 & $0.0018$  &   4882.76777  $^{+ 0.00023  }_{- 0.00023}$ &                                  33   &            33    \\     
992& $0.0016$  &   4885.63248  $^{+ 0.00029  }_{- 0.00030}$ &                                   12         &      13    \\     
1006  & $0.0017$   &     4905.68710  $^{+ 0.00033  }_{- 0.00032}$ &                                10         &      10   \\     
1045  & $0.0017$   &  4961.55291  $^{+ 0.00017  }_{- 0.00017}$ &                                -53       &      -52   \\      
1471& $0.0013$    &     5571.78736  $^{+ 0.00027}_{- 0.00026}$ &                                -46      &       -45    \\     
2530& $0.0034$    &     7088.77895  $^{+ 0.00056}_{- 0.00058}$ &                                -3        &       4     \\     
2585& $0.0025$    &     7167.56574  $^{+ 0.00058}_{- 0.00058}$ &                                54      &        62     \\     
\hline
\hline
\end{tabular}
\caption[caption]{\label{tab:resultados1} Central times of the
  transits of \ogle ~obtained from the light curve fitting
  with TAP and its residuals from the timing analysis. }
\end{table}

\begin{table}
\begin{tabular}{llll}
\hline
\hline
{ \scriptsize \textbf{Exp. Midtime} } &  {\scriptsize \textbf{Normalized}} & { \scriptsize  \textbf{Modelled} } & { \scriptsize   \textbf{Residuals} }\\
{ \scriptsize \textbf{($BJD_{TDB}$)} } &  {\scriptsize \textbf{Raw Flux}} & { \scriptsize  \textbf{Flux} } &  \\
\hline
$E=192$&&& \\
\hline
\small 
2453739.605173 &1.003039 &1.001438 &0.001601 \\
2453739.606043 &1.001873 &1.001420 &0.000453  \\
2453739.609924 &1.002305 &1.001340 &0.000965  \\
$...$ & $...$ & $...$ & $...$ \\

\hline
\end{tabular}
\caption{ \label{tab:LCS} Raw light curves of the nine transits of
  \ogle~presented in this work. We also included the best fitted
  model with TAP.   Full table is available in the online journal.  }
\end{table}

\section{Transit Timing Analysis} \label{analisis}

To ensure a uniform timing analysis for all transits, we converted the
time stamp in each new light curve frame to Barycentric Julian Days in
the Barycentric Dynamical Time standard system ($BJD_{TDB}$), as
suggested by \citet{Eastman_2010}, before modeling the light
curves. For the literature light curves we used the times provided by
A10, already converted to $BJD_{TDB}$.

We derived an $Observed$ $minus$ $Calculated$ $(O-C)$~diagram for the
19 modelled transits and the midtime reported by \citet{Konacki_2004}
for the transit on $E=-795$, using the constant period ephemeris
equation from A10, which has the form:

\begin{equation}
T_c = T_0 [BJD_{TDB}] + P*E,
\end{equation}
where $T_c$ is the predicted central time of transit in a given epoch
$E$, $T_0$ is the reference time of transit, and $P$ the orbital
period. The values of $T_0$ and $P$ adopted in this case are
$T_{0}=2453464.61762~BJD_{TDB}$ and $P=1.43247425$~days.  

\begin{figure}
\begin{center}
\includegraphics[width=1.\columnwidth]{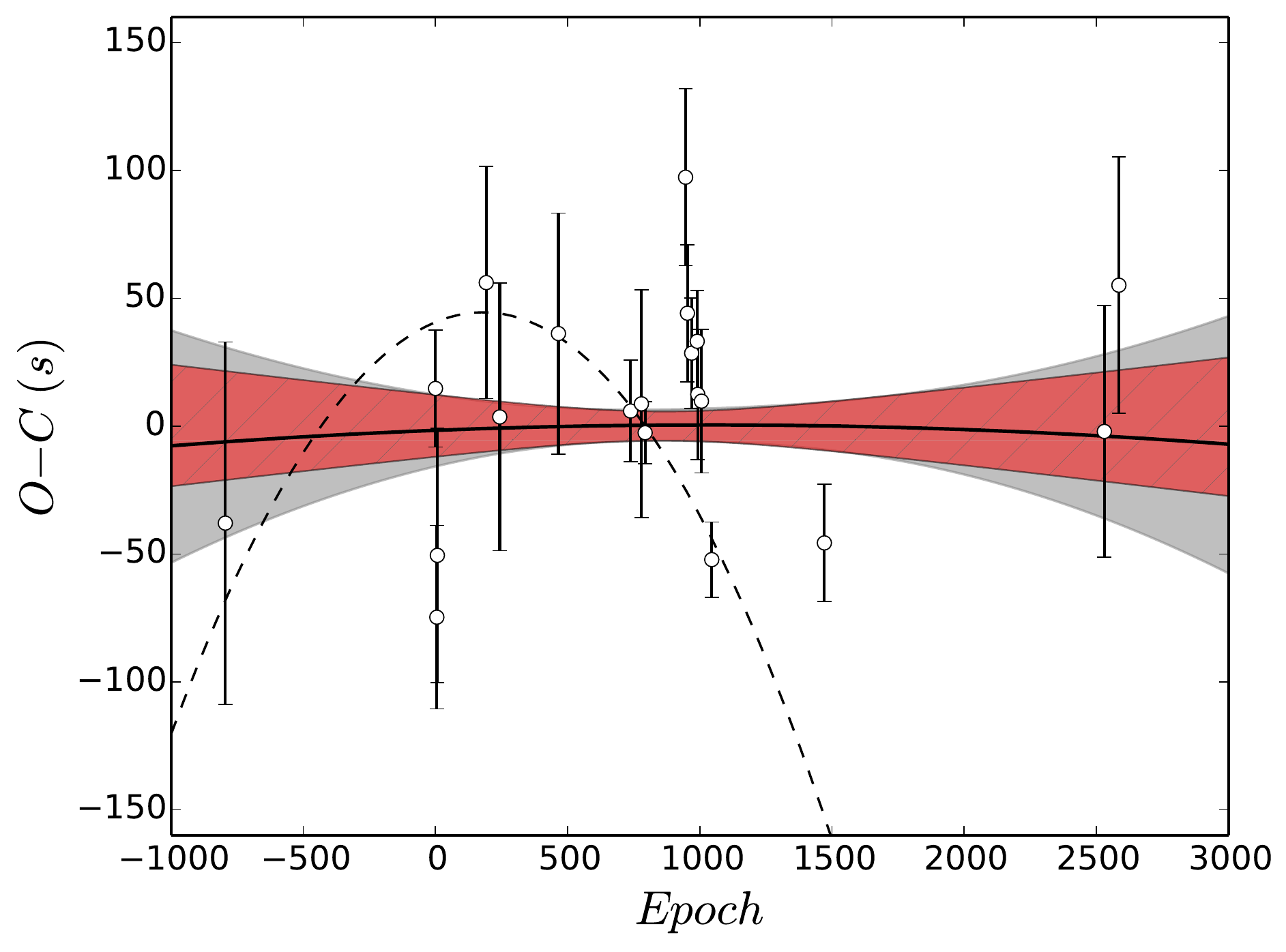}
\caption[]{\label{fig:oc1} Updated Observed \emph{minus} Calculated
  diagram for the transit midtimes of OGLE-TR-113b using a constant
  period ephemeris (eq. \ref{eqn:lineal1}).  The $\pm 1 \sigma$ errors
  of the linear ephemeris is represented by the red-hatched region.
  The RMS of the residuals is 42 seconds.  The linearly changing
  period ephemeris (eq. \ref{eqn:quadratic1}) is represented by the
  solid black line and its $\pm 1 \sigma$ errors by the gray region.
  The dashed line represents the corrected changing period function
  reported by Adams et al. 2010.}
\end{center}
\end{figure}

It is clear that the central times of the 20 transits do not follow
this ephemeris which can be due to accumulated uncertainty over time
on the parameters of the fit.  Therefore, in an attempt to correct for
those accumulated uncertainties, we perform a new weighted linear fit
to the transit midtimes.   This correction yields the following new
ephemeris equation:

\begin{equation}
T_c = 2453464.61708(14) [BJD_{TDB}] + 1.43247506(14) * E,
\label{eqn:lineal1}
\end{equation}
where the parameters and their $1 \sigma$ uncertainties are drawn from
their posterior probability distribution obtained from a MCMC analysis
performed with the \texttt{emcee} sampler implemented by
\citet{Foreman_emcee}. After correcting the ephemeris for this new
linear equation we obtain the timing residuals shown in the Figure
\ref{fig:oc1}.  The red-hatched region represents the $\pm1\sigma$
limits of the linear function. This new linear fit has a reduced
chi-squared of $\chi^{2}_{red}=2.3$ and a \textit{Bayesian Information
  Criterion} ($BIC$) of 48, while the dispersion of the timing
residuals is $RMS=42$~seconds.

Using the central times of 11 transits, A10 noticed a hint of an
orbital decay for \ogle~of $\dot{P}=-60\pm 15$~ms~yr$^{-1}$.  The
corrected version of the changing period function suggested by A10
(priv. communication) is represented by the dashed-line in Figure
\ref{fig:oc1}.  To check if this variation is still detected in our
extended dataset we fit our central times for the 20 transit epochs
for a linearly changing-period of the form (using the same notation
from A10):

\begin{equation}
T_c = T_0 [BJD_{TDB}] + P*E + \delta P * E(E-1)/2,
\label{eqn:quadratic1}
\end{equation}
where $\delta P$ represents the variation of the orbital period per
epoch ($P=P_0+\delta P*E$).  The quadratic fit is represented by solid
curve in Figure \ref{fig:oc1}.  Due to the low amplitude of the
quadratic term of the fit, the timing residuals of this fit are very
similar to the linear case.  The $ \pm 1 \sigma$ error of the
quadratic function is represented by the gray region of Figure
\ref{fig:oc1}.

We obtain a $\delta P =(-0.5 \pm 2.5) \times 10^{-10}$~days, which is
fully consistent with a constant orbital period ($\delta P = 0$) in
contrast with the value reported by A10 of $\delta P = (-2.74 \pm
0.66) \times 10 ^{-9}$~days. The dispersion of the midtimes residuals
of this quadratic fit is almost identical to the linear case
($RMS=41$~s) and with marginal differences in the statistical
indexes ($\chi^{2}_{red}=2.5$ and $BIC=51$).  In addition, when we
examine the change in period per year, we obtain $\dot{P}=-1.0 \pm
6.0$~ms~yr$^{-1}$, which is significantly smaller than the rate
observed before. 

As mentioned, the midtime of \citet{Konacki_2004} epoch is the
result of a combination of several low cadence light curves and
therefore is not well suited for timing analysis. Thus, we
explore the influence of this midtime in our ephemeris fits by
repeating our analysis without this epoch.  We observed no major
differences in the results of the weighted fits by excluding the
$E=-795$ transit, e.g., the quadratic term is consistent with zero ($
\delta P =( 0.2 \pm 2.8 ) \times 10^{-10}$ days).
   
Additionally, we find no evidence of periodic variations in the
timing residuals of the linear fit.  We also use the Anderson-Darling
test \citep{Anderson_1954} to probe if the residuals of the linear fit
are drawn from a Normal distribution.  According to this test, the
residuals sample comes from a normal distribution with 85$\%$ of
confidence.

Finally, we investigate the robustness of our results by exploring the
significance of the findings reported in A10 using our midtimes.  We
therefore, re-estimate equation \ref{eqn:quadratic1} using only our
values of $T_c$ from the literature transits, i.e. the light curves in
A10. We obtain a $\dot{P}=-44 \pm 21$~ms~yr$^{-1}$ which is smaller
but fully consistent with the value obtained by A10.  It is clear that
by adding our new transits in the O-C diagram, extending the system
monitoring time span from 6 to more than 13 years, the quadratic term
is much less significant than the one obtained with only the transits
up to 2009 ($E \sim 1000$).

\begin{table*}
\begin{tabular}{lcccccccc}
\hline
\hline
Fit & Period  & $T_0$ &  $\delta P$ & $\dot{P}$ & $\chi^{2}$ & $\chi^{2}_{red}$ & $BIC$ & $RMS$ \\
        & (days) & (BJD$_{TDB}$) & ($\times 10^{-10}$~days) & ($ms~yr^{-1}$) & &&& (seconds) \\
\hline
Linear & 1.43247506(14)  & 2453464.61708(14) & -- & -- & 42 & 2.3 & 48  & 42 \\
Quadratic   & 1.43247510(28) & 2453464.61706(16) & $ -0.5 \pm 2.5$ & $-1.0 \pm 6.0$  & 42 & 2.5 & 51 & 41 \\
\hline
\end{tabular}
\caption{ \label{tab:resultados4} Results of the linear and quadratic fits of the transit times of  \ogle. }
\end{table*}

\section{Perturber Mass Limits} \label{mercury}
\begin{figure}
\begin{center}
\includegraphics[width=1.\columnwidth]{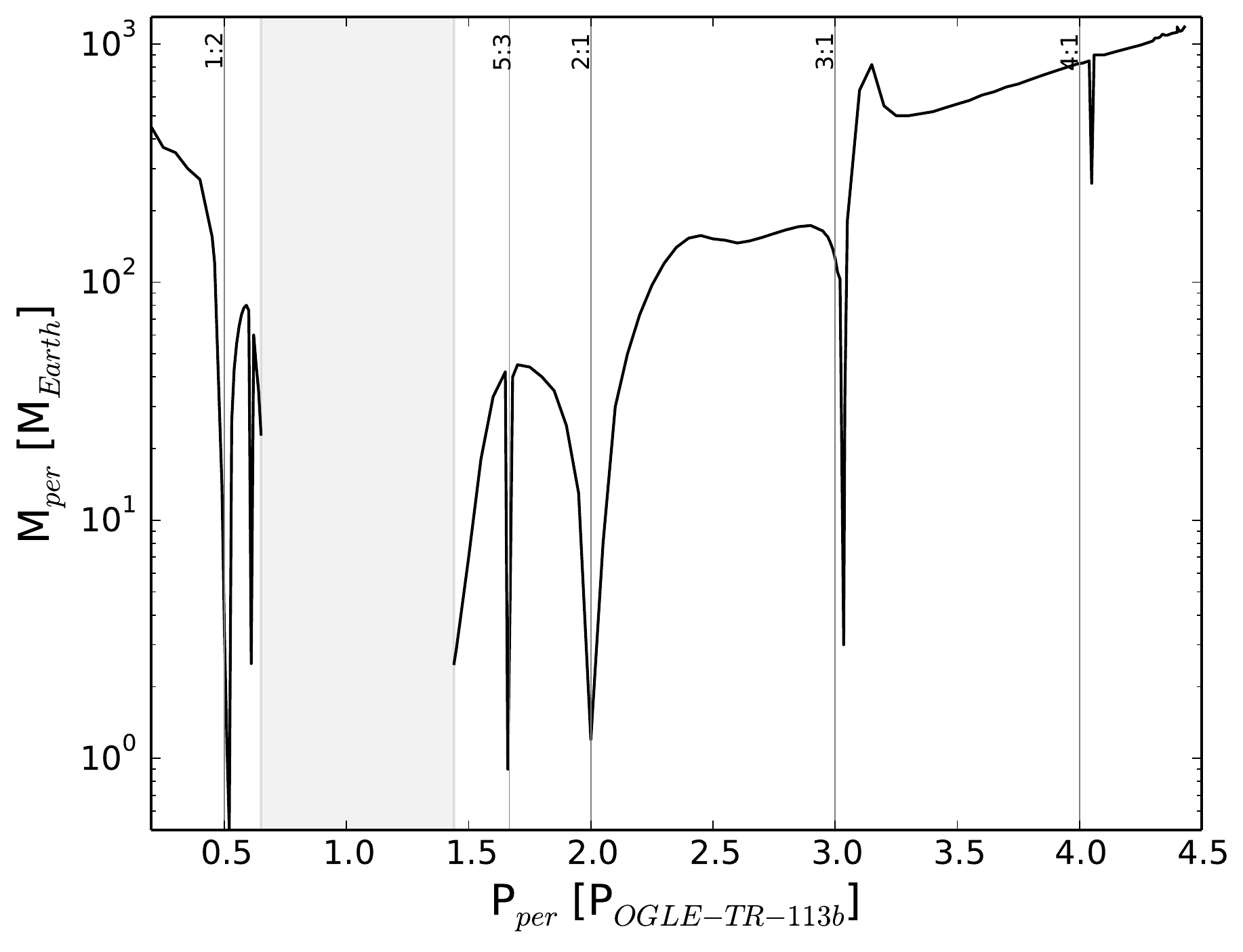}
\caption{\label{fig:mercury} Mass as a function of its orbital period
  of a hypothetical perturber in the OGLE-TR-113 system.  The gray
  strip and vertical lines represent the instability region of the
  system and the location of the principal Mean Motion Resonances,
  respectively.  }
\end{center}
\end{figure}

Using the limits imposed by our TTV analysis ($RMS=42$~seconds) we
investigate the mass of additional perturbing bodies in the system,
which could produce the observed dispersion in the transit midtimes.
For this, we use the \textit{Mercury} integrator code
\citep{Chambers_1999} to generate a set of dynamical simulations of
the OGLE-TR-113 system.  We use circular and coplanar orbits and set
the physical properties of the star and \ogle~to the values listed in
Table \ref{tab:resultados2}.  The initial orbit of the perturber was
calculated from Kepler's third law by using an orbital period in the
range $P_{per}= 0.1 - 4.5 P_{tran}$ in steps of $0.05$~or $0.005~
P_{tran}$ when more resolution was needed, e.g. near Mean Motion
Resonances (MMRs). $P_{tran}$ is the \ogle~orbital period derived in
this work.  The perturber mass varied from $0.1$ to
$1500$~$M_{\oplus}$; this variation depends on the calculated TTV (see
below).  We let the system evolve for 15 years but we save transit
times only after the first 3 years to avoid any perturbation induced
by initial conditions.  For each simulation we imposed the condition
that the calculated period of the \textit{transiting planet} did not
deviate more than $60$ seconds from the real period of \ogle.  If the
deviation was larger then the initial conditions of the transiting
planet's orbit for that specific simulation were changed in order to
obtain the desired orbital period.  Usually small changes in the
initial location of the planet were necessary.  Then, for each
simulation the $RMS$ of the TTVs was calculated, increasing the
perturber mass until an $RMS=45$~seconds was reached.  Close to this
mass level, we ran again the simulations using a mass step of $0.1$ or
$1.0$ $M_{\oplus}$, depending on the required precision.  The results
of these dynamical simulations are shown in Figure \ref{fig:mercury}.
By using the limits of our timing analysis we discard perturbers with
masses larger than 0.5 and 0.9 $M_{\oplus}$ near the 1:2 and 5:3 MMRs, 1.2
$M_{\oplus}$ near the 2:1 MMR and 3.0 $M_{\oplus}$ near the 3:1
MMR. While we agree with the mass limits placed by A10 in the 1:2 and
2:1 MMRs, our 5:3 and 3:1 MMR limits are almost one order of magnitude
more strict.

\section{Conclusions} \label{conclusiones}

We have observed nine new transits of \ogle~as part of TraMoS
project extending the time span of the observations from 6 years to over 13
years.  By performing a simultaneous timing analysis of these
transits and literature transits we tested the tentative detection of
orbital period decay for this planet reported by \citet{adams2010}.

Our timing analysis of 20 transit epochs discards the presence of a linearly
changing period of \ogle.  We obtain a $\delta P=(-0.5 \pm 2.5) \times
10^{-10}$~days which is fully consistent with a constant orbital
period for \ogle.  Our updated $\dot{P} = -1.0 \pm 6.0$~ms~yr$^{-1}$
is about 1 order of magnitude smaller than the value reported by A10
and consistent with zero.

For a large sample of Kepler planet hosts, \citet{penev2012} set a
strong limit on the tidal quality factor of $Q_{\star}\geq 10^{7}$.
In the case of \ogle, using a $1\sigma$ value based on our measured
orbital decay, i.e.  $\dot{P} = -7.0$~ms~yr$^{-1}$, stellar and
planetary masses from \citet{southworth2012}, and eqs. 5 and 7 from
\citet{Birkby2014} we obtain $Q_{\star}\sim 2.6 \times 10^{4}$ for
this sytem.  Those values of $Q_{\star}$~imply a $T_{shift}=
157$~seconds after 13.2 years, which is clearly not observed in the $O-C$
diagram in Figure \ref{fig:oc1}.  Using $\dot{P}=-1.0$~ms~yr$^{-1}$,
we obtain $Q_{\star}\sim1.8 \times 10^{5}$ and a $T_{shift}$ of
$22$~seconds, which is fully consistent with the RMS of the timing
residuals.  Therefore, based on our timing analysis we can discard
$Q_{\star}<10^{5}$.  A time shift of 100 seconds is expected in 7 more
years (i.e. in a total of 20 years of monitoring) if
$Q_{\star}\sim10^{5}$ and $\dot{P}$ is of only a few
ms~yr$^{-1}$. Only a 10 seconds shift is expected if
$Q_{\star}\sim10^{6}$ instead.  
Additionally, based also on the timing
analysis of the transits, we can place strict constraints on the mass
of additional bodies in the system.  We discard planets with masses
larger than 0.5, 0.9, 1.2 and 3.0 $M_{\oplus}$ near the 1:2, 5:3, 2:1
and 3:1 MMRs. Finally, with the homogeneous analysis of these data and
the literature transits, we update the physical properties of this
system.

\section{Acknowledgements}

We thank the anonymous referee for the useful comments that helped to
improve the quality of the manuscript and R. Alonso for helpful
discussion.

Based on observations obtained at the Gemini Observatory, which is
operated by the Association of Universities for Research in Astronomy,
Inc., under a cooperative agreement with the NSF on behalf of the
Gemini partnership: the National Science Foundation (United States),
the National Research Council (Canada), CONICYT (Chile), the
Australian Research Council (Australia), Minist\'{e}rio da
Ci\^{e}ncia, Tecnologia e Inova\c{c}\~{a}o (Brazil) and Ministerio de
Ciencia, Tecnolog\'{i}a e Innovaci\'{o}n Productiva (Argentina).

Based on observations obtained at the Southern Astrophysical Research
(SOAR) telescope, which is a joint project of the Minist\'{e}rio da
Ci\^{e}ncia, Tecnologia, e Inova\c{c}\~{a}o (MCTI) da Rep\'{u}blica
Federativa do Brasil, the U.S. National Optical Astronomy Observatory
(NOAO), the University of North Carolina at Chapel Hill (UNC), and
Michigan State University (MSU).

S.H. acknowledges financial support from the Spanish Ministry of
Economy and Competitiveness (MINECO) under the 2011 \emph{Severo
  Ochoa} Program SEV-2011-0187.  PR acknowledges Fondecyt \#1120299,
Anillo ACT1120.  DM is supported by the Millennium Institute of
Astropysics MAS from the Ministry of Economy ICM grant P07-021-F.  PR
and DM are also supported by the BASAL CATA Center for Astrophysics
and Associated Technologies PFB-06.

\bibliographystyle{mn2ev2}
\bibliography{refs}

\end{document}